\documentclass[12pt,preprint]{aastex}

\pdfoutput=1

\def\and{$\&$ }

\usepackage{wrapfig}

\begin{document}

\title{Flickering of 1.3 cm Sources in Sgr B2: \\Towards a Solution to the Ultracompact H~{\sc ii} Region Lifetime Problem}

\author{C. G. De Pree\altaffilmark{1}, T. Peters\altaffilmark{2}, M.-M. Mac Low\altaffilmark{3}, D. J. Wilner\altaffilmark{4}, W. M. Goss\altaffilmark{5}, R. Galv\'an-Madrid\altaffilmark{6}, E. R. Keto\altaffilmark{4}, R. S. Klessen\altaffilmark{7} \& A. Monsrud\altaffilmark{1}}

\altaffiltext{1}{Agnes Scott College, 141 E. College Ave., Decatur, GA 30030}
\altaffiltext{2}{Institut f\"{u}r Theoretische Physik, Universit\"{a}t Z\"{u}rich, Z\"{u}rich, Switzerland}
\altaffiltext{3}{American Museum of Natural History, New York, NY}
\altaffiltext{4}{Harvard-Smithsonian CfA, Cambridge, MA}
\altaffiltext{5}{National Radio Astronomy Observatory}
\altaffiltext{6}{European Southern Observatory, Karl-Schwarzschild-Str. 2, D-85748 Garching, Germany}
\altaffiltext{7}{Universit\"{a}t Heidelberg, Zentrum f\"{u}r Astronomie, Institut f\"{u}r Theoretische Astrophysik, Albert-Ueberle-Str. 2, 69120 Heidelberg, Germany}

\begin{abstract}
Accretion flows onto massive stars must transfer mass so quickly that they are themselves gravitationally unstable, forming dense clumps and filaments. These density perturbations interact with young massive stars, emitting ionizing radiation, alternately exposing and confining their H~{\sc ii} regions. As a result, the H~{\sc ii} regions are predicted to flicker in flux density over periods of decades to centuries rather than increasing monotonically in size as predicted by simple Spitzer solutions.
We have recently observed the Sgr B2 region at 1.3 cm with the VLA in its three hybrid configurations (DnC, CnB and BnA) at a resolution of $\sim$0\farcs25. These observations were made to compare in detail with matched continuum observations from 1989. At 0\farcs25 resolution, Sgr B2 contains 41 UC H~{\sc ii} regions, 6 of which are hypercompact. The new observations of Sgr B2  allow comparison of relative peak flux densites for the H~{\sc ii} regions in Sgr B2 over a 23 year time baseline (1989-2012) in one of the most source-rich massive star forming regions in the Milky Way. The new 1.3 cm continuum images indicate that four of the 41 UC H~{\sc ii} regions exhibit significant changes in their peak flux density, with one source (K3) dropping in peak flux density, and the other 3 sources (F10.303, F1 and F3) increasing in peak flux density. The results are consistent with statistical predictions from simulations of high mass star formation, suggesting that they offer a solution to the lifetime problem for ultracompact H~{\sc ii} regions.
\end{abstract}

\section{INTRODUCTION}
Dreher \& Welch (1981) first pointed out that the number of Ultracompact (UC) H~{\sc ii} regions in the Galaxy far exceeds expectations if simple free expansion at the thermal sound speed determines their sizes. Subsequent observations of Galactic UC H~{\sc ii} regions (e.g. Wood \& Churchwell 1989) only compounded this ``lifetime problem''. 
The Sgr B2 star forming region, highly extincted at optical and infrared wavelengths, lies near the Galactic Center. Among the most luminous star forming regions in the Galaxy, Sgr B2 is associated with a giant molecular cloud (GMC) with a mass of $\sim$10$^6$ M$_{\odot}$ . This region contains 49 H~{\sc ii} regions, 41 of which are ultracompact (d$<$0.1 pc, EM$>$10$^7$ pc cm$^{-6}$) and hypercompact (HC, d$< $0.05 pc, EM$>$10$^9$ pc cm$^{-6}$) H~{\sc ii} regions (Gaume et al. 1995).\footnote{There are several definitions of HC H~{\sc ii} regions; here we use the criteria of Kurtz 2002}  The multiplicity of sources makes it an ideal laboratory for testing theories of UC H~{\sc ii} region evolution. 

Peters et al. (2010a) have carried out high-resolution numerical simulations of star formation that account for heating from both ionizing and non-ionizing radiation. These simulations show that the accretion flows needed to form massive stars rapidly become gravitationally unstable. The orbits of newly born massive stars through the resulting dense clumps and filaments confine and expose their ionizing radiation (Peters et al. 2010a). Throughout the main accretion phase, the resulting H~{\sc ii} regions ``flicker''  between HC and UC sizes, and do not monotonically expand. The models of Peters et al. (2010b) show that as these ÒflickeringÓ UC H~{\sc ii} regions expand and contract, they take on the shapes defined by the morphological classifications of Wood \& Churchwell (1989), Kurtz et al. (1994) and De Pree et al. (2005).

The results of this model offer a solution to the UC H~{\sc ii} region lifetime problem, because accretion goes on more than ten times longer than the free-expansion time for an H~{\sc ii} region (Peters et al. 2010b).  The model predicts that UC H~{\sc ii} regions can experience scale length and flux density variations of $\sim$5\% per year. This result is based on a simulation of a collapsing core much smaller than Sgr B2, but the basic physics of gravitational instability leading to
flickering remains applicable. Such fluctuations have been seen in a few sources with multi-epoch VLA observations (e.g. Cep A, Hughes 1988; NGC 7538 IRS1, Franco-Hernandez \& Rodriguez 2004; G24.78+0.08, Galv\'an-Madrid et al. 2008). 

Sgr B2 has been extensively studied at radio wavelengths (Gaume et al. 1995, De Pree et al. 1998, Qin et al. 2011, De Pree et al. 2011). Gaume et al. (1995) published the first high resolution ($\theta_{beam}$=0\farcs25, $\sim$2000 AU) radio images of the Sgr B2 Main, South, and North star forming regions taken in 1989. The original 1.3 cm Very Large Array (VLA) continuum images were followed by (1) H66$\alpha$ (1.3 cm) radio recombination line (RRL) observations at the same resolution, (2) lower resolution ($\theta_{beam}$ = 2\farcs5) H52$\alpha$ (7 mm) RRL observations (De Pree et al. 1996), (3) high resolution ($\theta_{beam}$ = 0\farcs065, $\sim$600 AU) 7 mm continuum observations (De Pree et al. 1998) and (4) high resolution H52$\alpha$ RRL observations (De Pree et al. 2011).  The region has not been imaged with a full synthesis observation (all three hybrid arrays) since the original 1989 observations. In an effort to detect the predicted fluctuations, we have observed the Sgr B2 region 23 years after the original three hybrid configuration observations were made. We have detected significant changes in four of the 41 UC
 H~{\sc ii} regions, one in Sgr B2 North (K3) and three in Sgr B2 Main (F10.303, F1 and F3). Three of the four fluctuating sources are consistent with the definition of HC H~{\sc ii} regions.

\section{OBSERVATIONS and DATA REDUCTION}
The three new hybrid array observations were made in January 2012 (DnC), May 2012 (CnB) and September 2012 (BnA). Each observation was for a total of four hours (source and calibrators). Observations were made at 22.36 GHz and 20.46 GHz, separated into 16 IFs (intermediate frequencies), 8 for each RRL (H66$\alpha$ and H68$\alpha$). Observations were made in 32 MHz sub bands, with 128 channels (Open Shared Risk Observing - OSRO dual polarization), giving 250 kHz channels. With 8x2 contiguous 32 MHz channels (in each RRL), there was a total of 512 MHz bandwidth in the continuum. The H66$\alpha$ line was centered in IF 13, and the H68$\alpha$ line was centered in IF 5. The primary flux density calibrator for the observations was J1331+305. The phase calibrator was J1733-130, which also served as the bandpass calibrator. 

The continuum data sets were reduced using the Astronomical Image Processing System (AIPS) taking into account Appendix E in the AIPS Cookbook, which describes specific issues related to the reduction and analysis of VLA data after the VLA upgrade. Each of the datasets (DnC, CnB and BnA) were independently flagged, calibrated, imaged and then self calibrated. The calibrated data sets were then combined and self calibrated again (DnA with CnB, and then this pair with the remaining BnA data). The resulting image (referred to as the 2012 image) had a restoring beam size of 0.24\arcsec$\times$0.17\arcsec~and an rms noise of 0.16 mJy/beam. Since the recombination line emission was centered in specific channels in IF 13 (for the 22.36 GHz data), the line-bearing channels in this IF were excluded from the construction of the continuum image in order to avoid line contamination. The 2012 continuum image is shown in Fig. 1. These data will be presented more completely in the near future.

The 1.3 cm continuum image from Gaume (1995) was made from the two 12.5 MHz IFs  of those observations, so that the continuum emission was increased by the strength of the H66$\alpha$ line (an effect mentioned briefly in Gaume et al. 1995). Therefore, we have also newly reduced the 1989 observations in AIPS following Gaume et al. (1995), except that we constructed the continuum image only from a single 12.5 MHz IF, the one containing the weaker He66$\alpha$ line. As with the 2012 data, each of the datasets (DnC, CnB and BnA) were independently flagged, calibrated, imaged and self calibrated. The calibrated data sets were then combined and self calibrated again (DnA with CnB, and then this pair with the remaining BnA data). The resulting image (referred to as the 1989 image) had a restoring beam size of 0.29\arcsec$\times$0.22\arcsec~and an rms noise of 0.64 mJy/beam. The 2012 image was convolved to the slightly larger restoring beam of the 1989 image so that they both would have the same beam size.

We then made primary beam corrections to the two images using the AIPS task PBCOR. In order to normalize the flux density scale between the two images, peak flux densities were measured for 10 of the bright central sources in the 2012 image, and compared to the peak flux density of the same 10 sources in the 1989 image. These ratios were averaged, and found to be 1.16$\pm$0.05. We multiplied the 2012 image by a factor of 1.16 over the entire image (using the AIPS task MATHS). These images were then the ones that were compared in order to search for any changes in peak flux density.

\section{RESULTS}
As a first estimate, peak flux density values were measured using the AIPS task JMFIT. The source peak flux density values in the 1989 and 2012 images are similar for almost all of the 49 sources in Sgr B2 Main, North, and South, with four exceptions whose peak flux density values that changed by more than 10 times the rms noise in the 1989 image. The sources that exceeded this 10$\sigma$ cutoff are Sgr~B2 Main F10.303, F1 and F3, and Sgr B2 North K3. Changes between the two epochs are: F10.303 (16\% peak flux density increase, from 0.077 Jy beam$^{-1}$ to 0.090 Jy beam$^{-1}$), F1 (7\% peak flux density increase, from 0.153 Jy beam$^{-1}$ to 0.164 Jy beam$^{-1}$), F3 (5\% peak flux density increase, from 0.227 Jy beam$^{-1}$ to 0.239 Jy beam$^{-1}$), and K3 (20\% peak flux density decrease, from 0.107 Jy beam$^{-1}$ to 0.085 Jy beam$^{-1}$).  The ultracompact source F10.37 fell just below the 10$\sigma$ cutoff level, with the remainder of the ultra compact sources in Sgr B2 Main and North showing changes at the level of 1-5 times the rms noise in the 1989 image.

We compared the two images by aligning them with the AIPS task HGEOM, and then taking the difference between the images using the AIPS task COMB. The results of these comparisons are shown in Fig. 2 and Fig. 3. Fig. 2 shows the 2012 data in Sgr B2 North (grayscale), overlaid with the difference between the 2012 and the 1989 images (contours). The first negative contour is at the 10$\sigma$ level in the difference, and successive contours are indicated in the figure caption. Fig. 3 shows the 2012 data in Sgr B2 Main (greyscale), overlaid with the difference between the 2012 and the 1989 images (contours). The positive contours show the regions where there is more emission in the 2012 image than the 1989 image. First contours (positive and negative) are at the 10$\sigma$ level in the difference, and successive positive contours are indicated in the figure caption. The negative (dashed) contours to the SE of the F2 and F4 sources indicate the presence of some extended emission in the 1989 image associated with these two sources not detected in the 2012 image. The spur of emission to the west of F1 in Fig. 3 is associated with a faint source detected in the 2012 observations that is at the sensitivity limit of the 1989 observations. These difference images indicate that the F10.303, F1, F3, and K3 regions are the sources with significant ($>$10$\sigma$) changes in their peak flux density. 
\begin{figure}[t]
\begin{center}
    \includegraphics[width=0.85\textwidth]{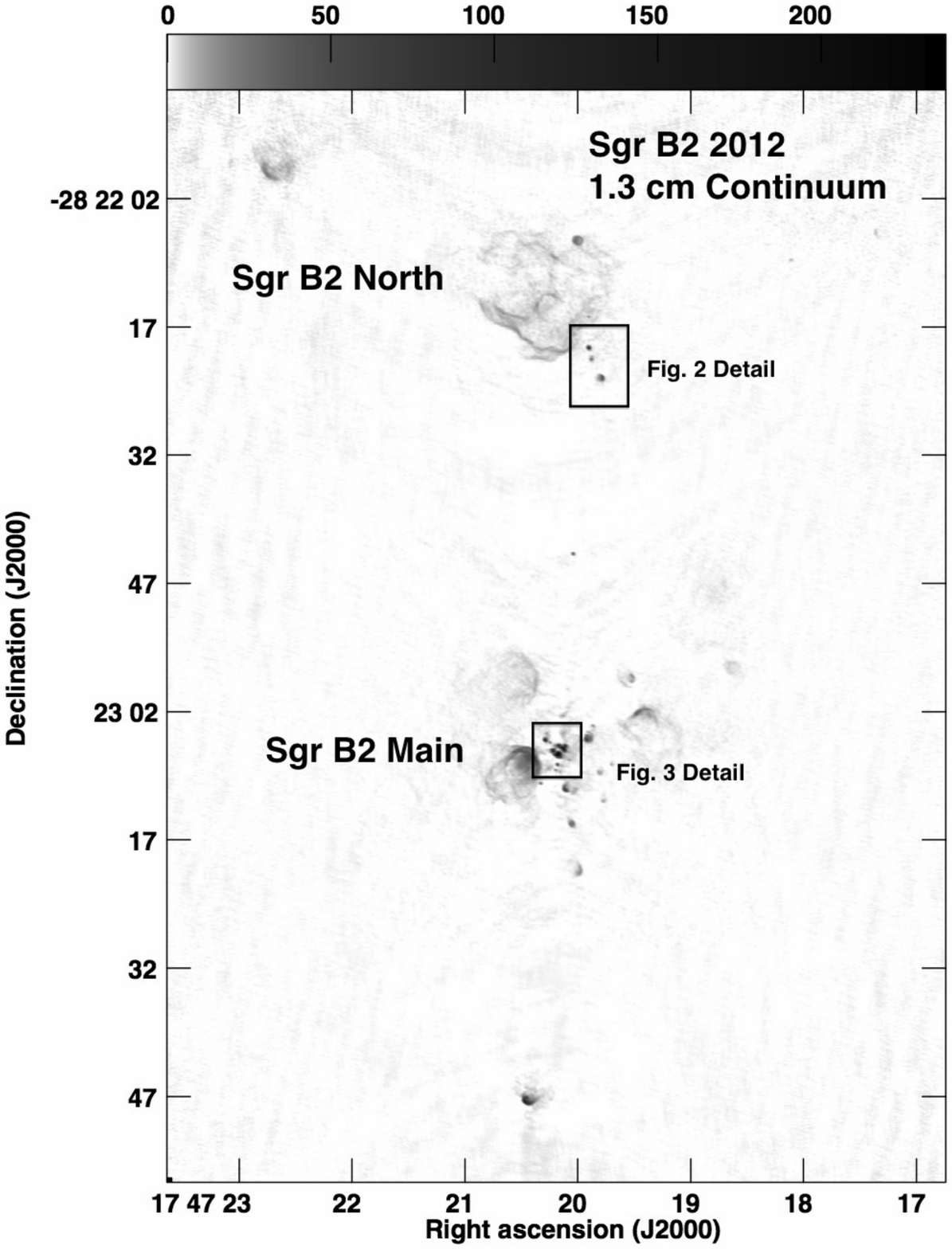}
  \end{center}
\caption{The 1.3 cm (22.4 GHz) 2012 VLA continuum image for the Sgr B2 region is shown in greyscale. Greyscale range is from 0 to 238 mJy beam$^{-1}$. The restoring beam in this image is $\theta_{beam}$ = 0.24\arcsec$\times$0.17\arcsec. The  rms noise is 0.16 mJy beam$^{-1}$  Regions of detail in Fig. 2 and Fig. 3 are indicated.}
\end{figure}

\begin{figure}[t]
\begin{center}
    \includegraphics[width=0.75\textwidth]{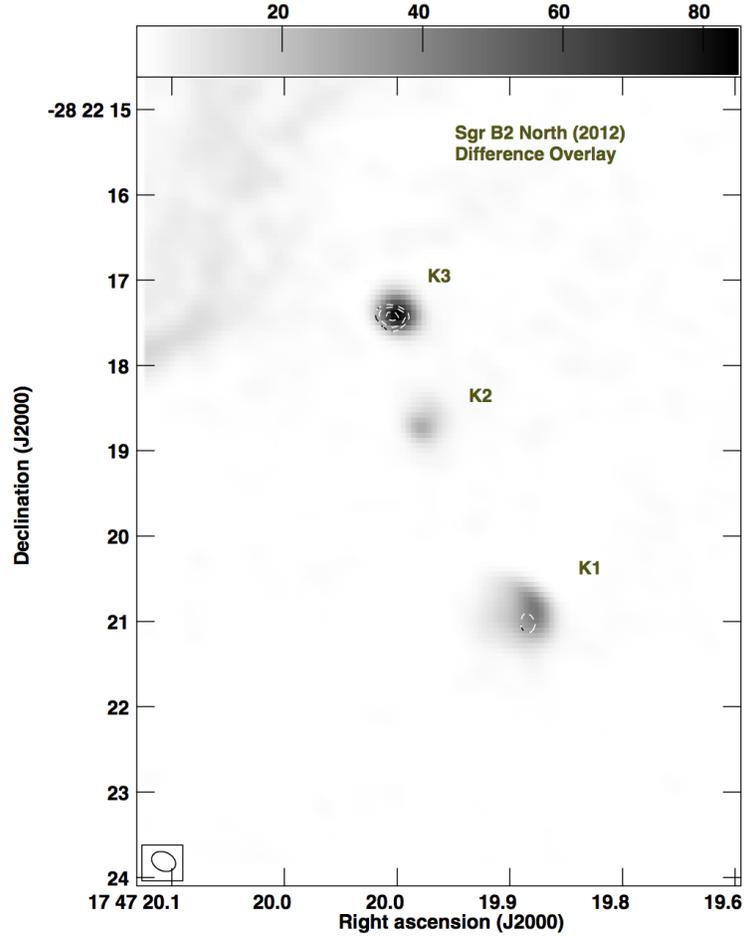}
  \end{center}
\caption{The 1.3 cm (22.4 GHz) continuum image for the central part of Sgr B2 North from the 2012 VLA data is shown in greyscale. The restoring beam in this image is $\theta_{beam}$ = 0\farcs29$\times$0\farcs22. Contours show the difference between the 2012 and the 1989 data in this field. First positive (solid) and negative (dashed) contours are at the 10$\sigma$ level. Successive negative contours are at 1.4 and 2 times the 10$\sigma$ contour.}
\end{figure}

\begin{figure}[t]
\begin{center}
    \includegraphics[width=0.85\textwidth]{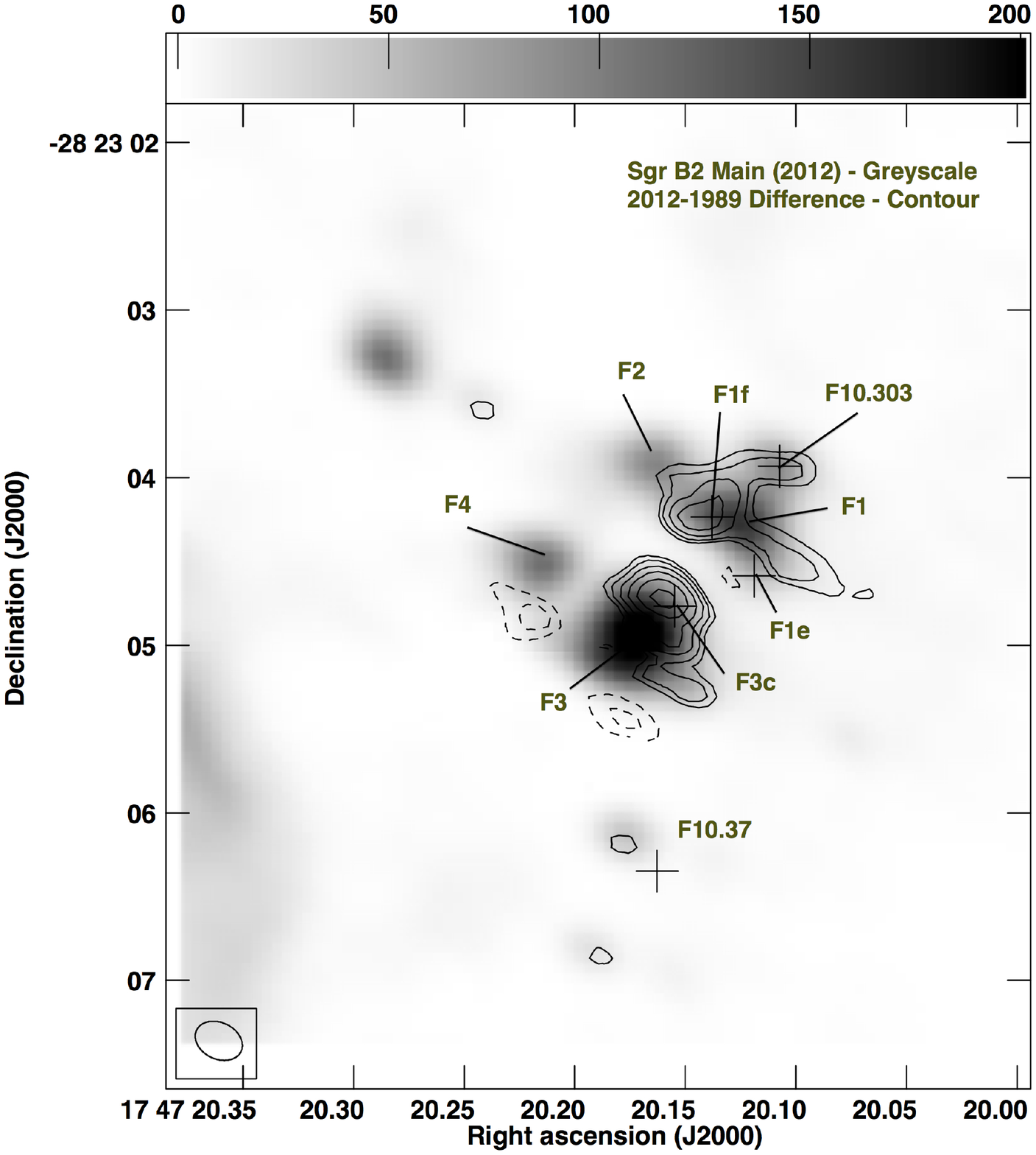}
  \end{center}
\caption{The 1.3 cm (22.4 GHz) continuum image for the central part of Sgr B2 Main from the 2012 VLA data is shown in greyscale. The restoring beam in this image is $\theta_{beam}$ = 0\farcs29$\times$0\farcs22. Contours show the difference between the 2012 and the 1989 data in this field. First positive (solid)  and negative (dashed) contours are at the 10$\sigma$ level. Successive positive and negative contours are at 1.4, 2, 2.8, 4 and 5.6 times the 10$\sigma$ contour. The positions of the five HC H~{\sc ii} regions with the highest electron density and emission measure (De Pree et al. 1998) are indicated with crosses and labeled.}
\end{figure}

\section{DISCUSSION \& CONCLUSIONS}
These continuum images represent the first results from observations of the continuum and radio recombination line emission at K band (1.3 cm) and Q band (7 mm) in the Galactic massive star forming region, Sgr B2 (Main, North, and South). This preliminary analysis indicates that four of the 41 UC H~{\sc ii}  regions in Sgr B2 Main and North have undergone a significant ($>$10$\sigma$) change in their peak flux density between matched resolution, full synthesis, hybrid configuration observations made in 1989 and 2012. The sources with flux density changes are F10.303, F1, F3 and K3. 


The frequency and scale variations are consistent with the
predictions of Galv\'an-Madrid et al. (2011). Their statistical
analysis is based on two collapse simulations with ionization
feedback from Peters et al. (2010a). Run A is a
numerical experiment in which only the central star was allowed to form, with subsequent star formation artificially
suppressed, while in Run B star formation was allowed to occur unimpeded, resulting in the formation of an entire cluster with three
$\sim$20 M$_\odot$~stars. The motivation for Run A was to investigate whether
ionization feedback can become strong enough to stop accretion (it cannot),
but locally it may be a reasonable model for stronger accretion flows than
those obtained in Run B, where fragmentation-induced starvation weakens
the infall onto the H II regions (Peters et al. 2010a, 2010c).

To create predictions for H II region variability on short time scales,
G\'alvan-Madrid et al. (2011) resimulated four flickering events from each
of these runs with outputs at $\sim$10 yr time resolution. 
In this limited sample, they find a probability of flux density increments exceeding 10\% over a 20 year baseline to be 21\% in Run A and 9\% in Run B, while the
probability of flux density decrements of the same magnitude was 7\% in Run A and 5\% in Run B. They averaged these results, somewhat arbitrarily, to yield the final quoted numbers of $\sim$15\% probability of a flux density increment and $\sim$6\% of a flux density decrement over 20 years.
None of these simulations represents a realistic model of Sgr B2
globally, nor were the flickering events chosen representative in any way.
In fact, the analysis of the global time evolution presented in
Galv\'an-Madrid et al. (2011) displays different phases of stronger and weaker
flickering activity, depending on the instantaneous infall rate
onto the UC H~{\sc ii} region, the geometry of the accretion flow, and the size
of the UC H~{\sc ii} region. Based on the models, and taking into account statistical errors at the 
2-3\% level, we would expect to observe 4-7 of the UC sources brighten (we detect 3), and 1-4 of the UC sources dim (we detect 1). 
We note finally that the statistical predictions of Galvan-Madrid et al. (2011)
are based on synthetic VLA observations at 2 cm, so that we
expect a slightly smaller effect in our 1.3 cm observations
because of the reduced optical depth. We consider the agreement of the quantitative predictions for time variability
of  less than a factor of two to be a notable success of the model, which could not have been
expected beforehand. Classical Spitzer evolution would have predicted no flickering at all.

Furthermore, the statistical analysis predicts that positive variations
are more likely to happen than negative ones, in agreement with our
findings in Sgr B2. The simulations also predict that flux density
decrements should occur more rapidly than flux density increments.
The UC H~{\sc ii}  regions shrink faster than they re-expand. The flux density decrement in K3 is larger in magnitude
than the flux density increments in F10.303, F1 and F3.

In our model, this behavior is due to the shielding of the ionizing source by its own accretion flow. When this happens, the UC H~{\sc ii} region shrinks on the recombination timescale. Positive changes, on the other hand are related to the re-expansion of the UC H~{\sc ii} region after the accretion event. Depending on the geometry of the accretion flow, the ambient density can vary in different directions. While the UC H~{\sc ii} region can expand very quickly as an R-type front in under-dense regions, the expansion into a dense medium as a D-type front happens on much longer hydrodynamical timescales. We thus expect positive variations to be smaller on average than negative ones. 

D-type fronts cannot grow much faster than the sound speed, and thus
the UC H~{\sc ii} region expansion is limited to $\sim$100 AU in $\sim$20 yr. R-type fronts,
on the other hand, reach the Str\"omgren radius within a few recombination
times ($\sim$10 yr for a UC H~{\sc ii} region). Since size and flux density variations are
tightly correlated (Peters et al. 2010a), this may suggest that the large
flux density increments observed in the F sources are caused by R-type expansion
into a pre-existing cavity, which may have been created by the ionizing
radiation in a previous expansion phase (see the online material of
Peters et al. 2012 for three-dimensional visualizations of this
process).

The shrinkage of the H~{\sc ii} region K3 is evidence for accretion onto the massive star that powers K3. As pointed out by Franco-Hernandez \& Rodriguez (2004), such a drop in flux density can only be caused by intrinsic changes in the powering source or by
enhanced absorption of ionizing photons close to the star. Since the
timescale for structural changes of the protostar is much too large to
explain the observed sizeable flux decrement (Klassen, Peters \& Pudritz
2012), the increased absorption of ionizing radiation by the accretion
flow remains the only explanation. All four sources with changes in peak flux density have been observed in previous observations (De Pree et al. 1996, De Pree et al. 2011) to have broad radio recombination line (RRL) profiles (F10.303 \& F3 $\Delta$V$_{FWHM} >$ 50 km s$^{-1}$ in the H52$\alpha$ RRL and F1 \& K3 $\Delta$V$_{FWHM} >$ 35 km s$^{-1}$ in the H66$\alpha$ RRL). 
Analysis of the new H52$\alpha$, H66$\alpha$ and H68$\alpha$ RRL data could yield additional evidence for infall of gas through the H~{\sc ii} region (Peters, Longmore \& Dullemond 2012). 

We note that the three sources that have significantly brightened in Sgr B2 Main (F10.303, F1, and F3) were observed in the 7 mm continuum with 0\farcs065 resolution with the VLA (De Pree et al. 1998). In that 7 mm study, the five sources with the highest electron density and emission measure as determined from the 7 mm continuum (Table 2 in that paper) are F10.303 (F1a in that paper), F1e, F1f, F3c and F10.37. Crosses in Fig. 3 indicate the positions of these sources. All of these sub sources fall at or near the criteria for hypercompact sources. The F1f sub source is located on the eastern edge of F1 (where it has been observed to brighten), and the F3c sub source is at the northern edge of F3 (where its brightening is observed). The brightening in F10.303 (F1a) appears to be centered on the source. As mentioned above, F10.37 fell just below our 10$\sigma$ cutoff criterion for significant change.

These observations are consistent with the proposal that flickering can resolve the lifetime problem for UC H~{\sc ii} regions, as proposed by Peters et al. (2010b). They further lend support to those authors' model for massive star formation (Peters et al. 2010a, 2010c), in which gravitational instability and the accretion flow causes it to fragment into both dense filaments and the secondary stars that inevitably accompany massive stars, resulting in fragmentation-induced starvation of the most massive stars (see also Girichidis et al., 2012).

\acknowledgements{This research was supported by a grant from the National Science Foundation (AST-1211460). C. G. De Pree gratefully acknowledges J. Heath, M. Hutcheson, K. Luong and K. Butler who worked on aspects of this project. M.-M. Mac Low was partially supported by NSF grant AST-1109395. T.P. acknowledges financial support through SNF grant
200020\textunderscore 137896 and a Forschungskredit of the University of Z\"{u}rich, grant no. FK-13-112. R. S. Klessen acknowledges financial support by the German Science Foundation (DFG) in the collaborative research project SFB 881 ``The Milky Way System'' in sub projects B1, B2, B3, and B5.} We acknowledge the anonymous referee for a careful reading that improved the paper.

\end{document}